\documentclass{article}

\usepackage{arxiv}
\usepackage[utf8]{inputenc} 
\usepackage[T1]{fontenc}    
\usepackage{hyperref}       
\usepackage{url}            
\usepackage{booktabs}       
\usepackage{amsfonts}       
\usepackage{nicefrac}       
\usepackage{microtype}      
\usepackage{lipsum}		
\usepackage{graphicx}
\usepackage{natbib}
\usepackage{doi}
\usepackage{listings}
\usepackage[most]{tcolorbox}
\usepackage{tabularray}

\title{Harnessing the Power of LLMs: Automating Unit Test Generation for High-Performance Computing}

\date{October 9, 2023}	

\author{ \href{https://orcid.org/0000-0002-6705-6506}{\includegraphics[scale=0.06]{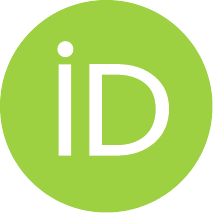}\hspace{1mm}Rabimba Karanjai}\thanks{www.rabimba.me} \\
	Department of Computer Science\\
	University Of Houston\\
	\texttt{rkaranjai@uh.edu} \\
        \And
	Aftab Hussain \\
 	Department of Computer Science\\
        University Of Houston \\
	\texttt{ahussain27@uh.edu}\\
         \And
	Md Rafiqul Islam Rabin \\
 	Department of Computer Science\\
        University Of Houston \\
	\texttt{mrabin@uh.edu}\\
        \And
	Lei Xu \\
 	Department of Computer Science\\
        Kent State University \\
	\texttt{xuleimath@gmail.com}\\
        \And
	Weidong Shi \\
 	Department of Computer Science\\
        University Of Houston \\
	\texttt{wshi3@uh.edu}\\
        \And
	Mohammad Amin Alipour \\
 	Department of Computer Science\\
        University Of Houston \\
	\texttt{maalipou@central.uh.edu}\\
}

\renewcommand{\shorttitle}{\textit{arXiv} Template}

\hypersetup{
pdftitle={A template for the arxiv style},
pdfsubject={q-bio.NC, q-bio.QM},
pdfauthor={David S.~Hippocampus, Elias D.~Striatum},
pdfkeywords={First keyword, Second keyword, More},
}

\begin{document}
\maketitle

\begin{abstract}
	Unit testing is a standard practice in software engineering and is critical for ensuring software quality. 
    However, for parallel and high-performance computing software, especially scientific computing applications, unit testing is not widely implemented. 
    Compared with typical commercial software, high performance software usually have a smaller user base, and they are diverse and usually involve complex logic.
    These characteristics create several challenges for conducting unit testing for parallel and high performance software.
    On one hand, it is economically expensive to have a dedicated testing team to do unit testing considering the number of users.
    On the other hand, it is hard for a quality engineer without domain knowledge to design effective unit testings. Similarly, existing automated unit testing tools are usually not effective for such software.
    Therefore, it is vital to devise an automated method for generating unit testing cases for parallel and high performance software, which considers the unique features of these software, including complex logic and sophisticated parallel processing techniques. 
    Recently, large language models (LLMs) have attracted more attention and are believed to be a powerful tool for coding and testing, but its application in producing unit tests for parallel and high performance applications remains uncertain. 
    To fill this gap, we explore the capabilities of two well-known generative models, Davinci (text-davinci-002) and ChatGPT (gpt-3.5-turbo), in crafting unit testing cases for parallel and high performance software. 
    Specifically, we proposed novel ways to utilize LLMs to develop unit testing cases for high performance software with \texttt{C++} parallel programs and assessed their effectiveness on extensive \texttt{OpenMP}/\texttt{MPI} projects. 
    Our findings indicate that in the context of parallel programming, LLMs can create unit testing cases that are mostly syntactically correct and offer substantial coverage, while they exhibit some limitations like repetitive assertions and blank test cases.
\end{abstract}


\section{Introduction}
Fueled by advancements in deep learning, Large Language Models (LLMs) have surged in development. Their versatility stretches across various applications, with a prominent one being software code generation. LLMs specializing in this domain are trained on vast datasets of source code gleaned from software repositories like GitHub and programming communities like Stack Overflow. OpenAI Codex~\cite{chen2021codex}, Meta's Code Llama (a code-centric variant of Llama 2 trained on dedicated code datasets), and HuggingFace's StarCoder~\cite{li2023starcoder} stand out as notable examples. Early research and evaluations suggest that these code-tailored LLMs exhibit impressive competency in generating programming code from user-provided natural language queries~\cite{chen2021codex, 10.1145/3520312.3534862}.

Incorporating Large Language Models (LLMs) into the realm of software development holds the potential to unlock remarkable benefits, spanning a diverse range of tasks, including code creation, documentation, summarization, analysis, and troubleshooting~\cite{tarassow2023potential}. The capabilities of LLMs extend to practical assistance in code restructuring, choosing suitable names for variables and functions, and identifying and removing redundant code sections. These advanced functionalities can streamline numerous daily tasks faced by software engineers. Moreover, the ability of LLMs to generate code offers the promise of automating various phases of the development process, leading to enhanced efficiency \cite{svyatkovskiy2020IntelliCode,github2021copilot}. Recognizing this potential, developers are swiftly embracing code-generating LLMs into their workflows. A noteworthy example is Copilot, powered by the Codex model, which has garnered a user base exceeding one million developers \cite{github1m}. GitHub reports that developers utilizing GitHub Copilot \cite{github2021copilot} have experienced a 55\% boost in development speed, with nearly 40\% of their code being generated by the model.

In the realm of software engineering, a domain ripe for LLM augmentation is unit testing, particularly within the fundamental practice of programmer-authored unit tests. These meticulously crafted tests scrutinize individual code segments, ensuring their functionality and acting as sentinels against lurking bugs. Their significance and benefits have been extensively documented in various studies~\cite{daka2014utsurvey,gren2017relation,10.1145/1159733.1159788,10.5555/1802408.1802420}. 
Furthermore, unit tests readily lend themselves to automation, becoming invaluable players in the software development's regression testing arsenal. Implementing such fine-grained regression testing bolsters both maintainability and dependability, serving as a crucial pillar for robust software projects~\cite{taneja2008diffgen}. In contemporary programming practices, unit tests command the same reverence as production code, adhering to stringent quality standards, undergoing rigorous checks, and residing within shared version control systems~\cite{tosun2018effectiveness}.

Large-scale software development teams wrestling with unit testing often face steep time commitments and formidable challenges, as Schafer's 2023 study vividly illustrates \cite{schafer2023empirical}. Developers become time-jugglers, balancing production code with crafting a plethora of tests to blanket their modules with adequate coverage. Frequently, the sheer volume of these test scripts eclipses the functional code. Enter the cavalry: automated unit test generation tools devised by researchers. These tech-powered knights wield diverse arsenals, ranging from fuzzing's exploratory blasts~\cite{miller1990empirical} to coverage analysis's strategic mapping~\cite{taneja2008diffgen}, and even symbolic execution's logic-defying maneuvers~\cite{tillmann2014transferring}. Search-based and evolutionary algorithms also join the fray, their battle cries echoing promises of lighter developer workloads and potent tests that illuminate code mysteries and bolster coverage~\cite{lampropoulos2019coverage}

There are however unique challenges for the methods used to generate automatic unit tests. For example, research has highlighted that tests produced automatically often lack the readability and understandability of those manually crafted by developers~\cite{almasi2017industrial,grano2018empirical}.
In an effort to overcome these limitations, there is an increasing interest in the application of machine learning, particularly Large Language Models (LLMs), to enhance both the process and output of unit test generation~\cite{mastropaolo2021studying, chen2022codet,lahiri2022interactive,bareiss2022code,schafer2023empirical}.
The underlying idea is that LLMs, trained on extensive corpuses of natural language and coding material, might be adept at creating more developer-friendly test code.
More specifically, these LLMs could be capable of creating unit tests that mirror those produced by human developers~\cite{schafer2023empirical}, thus potentially addressing the issues inherent in other automated unit test generation methods.
In this context, recent studies have focused on using LLMs for generating Java code tests~\cite{bareiss2022code, tufano2020unit}, demonstrating promising improvements over traditional methods like feedback-directed test generation~\cite{pacheco2007randoop}.

Most of the research so far applying LLMs for test generation focused on general-purpose programming tasks, and little attention has been paid to the feasibility of applying code-generating LLMs in the context of testing parallel programs or high-performance computing (HPC) programs.  
In general, testing parallel programming code and applications is considered to be challenging~\cite{4548405,thies2021rse}. Many large scale parallel computing machines or HPC installations are one of a kind, often facing unknown issues and difficulties to predict their behaviors. Such unique environment makes unit testing both important and challenging. For instance, after optimizing a piece of parallel high performance code, it is important to run well prepared regression tests to tell whether the code behavior has changed with the optimizations. 
In addition, HPC applications are parallel programs distributed across large clusters. The code is typically based on concurrent programming models such as MPI (Message Passing Interface) for distributed memory and OpenMP (Open Multi-Processing) for shared memory. These programming models support the concurrent execution of code, which introduces new challenges for testing such as deadlocks and race-conditions. Most existing test framework lacks support for HPC applications, and parallel programs~\cite{thies2021rse}.

In this paper, we investigate LLM-based software testing from another perspective, i.e., instead of considering the code written in a specific language, we focus on the type of programs that involve parallel computing, which usually contains more complex logic and are prone to have subtle errors, which makes them difficult to test. Specifically, we conduct an empirical study on LLM-aided unit test generation for C++ code utilizing OpenMP for high-performance computing. 
We utilize two LLMs, Davinci (text-davinci-002)~\cite{godoy2023evaluation} and ChatGPT (gpt-3.5-turbo)~\cite{dong2023self}, to generate unit tests for the existing code repositories and evaluate the quality by checking their usefulness in testing.

We addressed three specific research questions in the parallel programming domain. Firstly, we explored the use of the entire class under test as context for the LLMs to generate unit test cases. Secondly, we examined how different context styles, such as solely using the method under test, code compatibility, code quality, and other aspects. Thirdly, we explore if the generated test cases can test the hierarchical parallelism in the source codes.
This research question is particularly significant in the parallel programming domain as it focuses on generating unit tests for high performance code that employs OpenMP in C++. The ability to automatically generate effective unit tests for such programs can greatly enhance HPC software quality, performance, and reliability. We adopt the approach employed by Siddiq et al.~\cite{siddiq2023exploring}, who delved into the applicability of LLMs in the context of unit test cases. It's worth noting that their study focused on the HumanEval and EvoSuite datasets, but they did not specifically address parallel or high performance code.

The projects we choose, uses OpenMP and MPI for parallelization. The unit tests we generate test those programs. 
These are all regular C++ programs, but since they use these specific libraries what they test is different than normal cpp programs. That is our main focus. to see if the LLMs can catch these intricacies and generate unit tests for these high performance programs.
The contributions of our work can be summarized as the follows:

\begin{itemize}
    \item Our research entailed an exhaustive analysis of LLMs to produce unit tests for high performance and parallel programs in a zero-shot framework. A total of 216 unit tests were created from seven open-source high performance computing projects, examined across three distinct temperature settings, culminating in 648 test cases altogether.
    \item An methodical evaluation of the unit tests' quality generated by Davinci and ChatGPT in the parallel and high performance computing arena was conducted, focusing on the incidence of test smells.
    \item The study delved into the influence of context on code generation and assessed the impact of incorporating source code and function context in enhancing unit test creation for LLMs in the parallel and high performance programming field.
    \item A thorough discussion on the broader consequences of applying code generation models for crafting unit tests in high performance computing was presented.
\end{itemize}

\section{Motivation}

Large Language Models (LLMs) can struggle with understanding and writing OpenMP and MPI unit tests due to the inherent complexity and domain-specific nature of parallel programming. These frameworks introduce concepts like thread synchronization, data distribution, and inter-process communication, which are not easily captured by the general language patterns LLMs are trained on. Furthermore, OpenMP and MPI code often involves low-level system calls and platform-specific optimizations, posing additional challenges for LLMs trained primarily on high-level natural language data. This difficulty is reflected in research highlighting the limitations of LLMs in code generation tasks involving specialized libraries and frameworks \cite{chen2021evaluating,wang2023codet5+}.

Unit testing is crucial for OpenMP and MPI codes due to their inherent complexity and susceptibility to subtle errors arising from race conditions, deadlocks, and non-deterministic behavior \cite{wittwer2006introduction}.  Unlike non-high-performance codes, where errors often manifest as incorrect outputs, parallel programming errors can be silent and sporadic, making them difficult to diagnose and reproduce \cite{pachecoparallel}. Comprehensive unit tests provide a safety net by isolating individual code units, enabling developers to verify their correctness and pinpoint issues early in the development process \cite{gropp1999using}. Additionally, unit tests serve as regression tests, ensuring that code changes do not inadvertently introduce new errors or break existing functionality \cite{beck2022test}. This is especially important for OpenMP and MPI codes, which often undergo frequent modifications and optimizations to improve performance and scalability.

Given the challenges of manual test creation for OpenMP and MPI code, automated test generation emerges as a critical solution. Automated tools can systematically explore the vast parameter space of parallel programs, uncovering corner cases and potential errors that may be overlooked by human developers \cite{olsthoorn2022more,gopalakrishnanformal}. By generating diverse and comprehensive test suites, these tools can significantly enhance the reliability and robustness of parallel applications, reducing the risk of costly failures in production environments \cite{andrews1993sr}. Additionally, automated test generation can accelerate the development process by relieving developers from the tedious and error-prone task of manual test writing, allowing them to focus on higher-level design and optimization \cite{XieN2006}.

\section{Methodology}\label{sec-method}

\subsection{Data Collection}
The majority of the previous studies on code generation with the LLMs have primarily utilized the datasets from HumanEval~\cite{chen2021codex} and Humaneval X~\cite{zheng2023codegeex}. 
However, these two datasets are not suitable for our purpose because they lack parallel and high performance related C/C++ code that utilizes OpenMP.

To overcome this challenge, we thoroughly review a large number of renowned high performance computing projects from GitHub spanning various sub-fields and select a subset from them to serve as the benchmark of the newly developed LLM-based unit test generation. Our selection criteria are as follows:

\begin{itemize}
    \item The projects should be popular (e.g., starred/forked count) so that we can hypothesize that they have acceptable quality human-written code.
    \item The code repository should have well-documented unit tests written in English.
    \item The unit tests should all be standalone so that code generation does not suffer from dependency problems.
    \item Unit tests themselves should have a proper structure, so that while generating code using the LLMs, we can guide the generation process for better outcome.
\end{itemize}

Finally, 7 projects are chosen for data collection, which are summarized in \tablename~\ref{projects}.

\begin{table}[!ht]
    \centering
    \footnotesize
    \caption{Collected parallel and high performance programming projects along with unique test cases.}
    \label{projects}
    \begin{tabular}{ll}
        \toprule
        Project & Unit Tests \\
        \midrule
        Optim \cite{Goptim} & 21\\
        DBCSR \cite{dbcsr} & 6\\
        Arraymancer \cite{arraymancer} & 22\\
        CTranslate2\cite{c2} & 33\\
        FAASM\cite{faasm} & 39\\
        AMGCL\cite{amgcl} & 11\\
        Stats\cite{stats} & 84\\
        \bottomrule
    \end{tabular}
\end{table}

\subsection{Test Case Generation}

We investigated the following research questions to determine how well the large language models (LLMs) can generate unit tests for the parallel programming use cases. We followed the approach of Siddiq et al.~\cite{siddiq2023exploring}, who also explored the potential of using LLMs for unit test cases with the HumanEval and EvoSuite datasets. However, our focus is on the parallel and high performance domain. We explain our test case generation process in detail in Section \ref{testcgen}.
We primarily aim to answer the following research questions through our work.

\begin{tcolorbox}
    \textbf{Research Question 1}

    How effectively can LLMs generate unit tests for parallel and high performance programs?
\end{tcolorbox}

We choose two LLMs as the backend for unit test generation for the parallel and high performance projects listed in \tablename~\ref{projects}, Davinci (text-davinci-002)~\cite{godoy2023evaluation} and ChatGPT (gpt-3.5-turbo)~\cite{dong2023self,openai2022chatgpt}. Both are developed by OpenAI and corresponding APIs are used to interact with them.
For each test, we select 10 candidates with three different temperatures $0$, $0.2$, and $0.4$. 
Here \textit{temperature} is a hyper-parameter that regulates the randomness, or creativity, of the LLM’s responses.
Value $0$ will lead to more predictable results (i.e., the response is deterministic and less creative), and the other two temperatures are selected to explore potential positive impacts on unit test generation. It is possible that a higher temperature may lead to worse performance. The test case generation method and details on the different considerations are provided in Section~\ref{testcgen}.

\begin{tcolorbox}
    \textbf{Research Question 2}

    Does the template guided code generation improve the unit tests?
\end{tcolorbox}

Our assumption is that the programmers will employ specific prompts similar to ``Can you generate a unit test for a code'' or something along that line. Often these prompts will encompass a code function to be tested. These often are sent as ``Write a Unit Test to test <<some task>> functions as needed''.
Here, the function being tested serves as the backdrop for the directive. 
This backdrop, or context, can encompass various code components. In this study, we delve into the influence of these distinct components on the unit tests produced by controlling the dataset in different ways, including:
\begin{itemize}
    \item Providing the full code context.
    \item Omitting the context entirely.
    \item Solely presenting the associated libraries for guidance as a code template.
\end{itemize}

This approach may shed light on the efficacy of LLMs when steered by specific input cues. We gauged their efficiency by examining compilation success rates, the breadth of code coverage, and the aggregate of accurate unit tests produced.

\begin{tcolorbox}
    \textbf{Research Question 3}

    Do the generated Unit Tests also cover hierarchical parallelism checks for the original source code?
\end{tcolorbox}

Hierarchical parallelism refers to the organization and execution of parallel tasks in a structured and layered manner. In parallel computing, this approach allows for the efficient decomposition of complex programs into more manageable sub-tasks, which can then be executed concurrently across multiple levels of parallelism. OpenMP, a popular parallel programming model, supports the nesting of parallel regions \cite{NestingofRegions}, enabling developers to define parallel tasks within already parallelized sections of code. This nested parallelism is crucial for harnessing the full potential of modern multi-core and multi-threaded processors, especially when dealing with multifaceted computational problems. However, with the increased complexity introduced by nested regions, unit tests become paramount. These tests ensure each individual component or function operates correctly in isolation, helping to identify and rectify issues that might arise due to the intricacies of hierarchical parallelism.

We further evaluate if the unit tests that we have generated can check whether these parallelism implementations are present in the tested code or not, and whether they are implemented in the correct way.

\section{Test Case Generation with LLMs}\label{testcgen} 
In this section, we investigate three distinct methods for leveraging large language models (LLMs) in generating unit tests for parallel and high-performance computing programs. We assess template-based generation, where existing unit test code serves as a structural blueprint for the LLM; contextual generation, where the LLM is informed by the codebase context to create relevant tests; and guided generation, involving iterative interaction with the LLM through prompts and feedback. 

\subsection{Out of the Box Test Case Generation}
The out-of-the-box (OOB) approach is the most straightforward one to generate test cases for parallel and high performance programs. 
With the OOB approach, we do not provide any contextual information to the LLMs, and instead just rely on the test case patterns for code generation.
For most of cases, the unit test case generation can be treated as a special type of code generation problem with specific expectations. 
These expectations were later used to evaluate how good or bad the code generation task was completed in the context of unit tests.

For unit tests, each of the prompts that we used was derived by giving a candidate code to the API. Even though both the evaluated LLMs are capable of generating code, the number of tokens that they can support and the memory capability is different. Due to these differences, we used a slightly different approach for unit test generation applying the LLMs.

To access both models, we used OpenAI's APIs \footnote{https://platform.openai.com/docs/api-reference}. The APIs provide a convenient way to access and evaluate the models, as well as retrieve the outputs. For experiment orchestration, we used Langchain~\cite{Chase2022LangChain} to load the test cases into memories for the conversation with the OpenAI models. Each of the test case generations used a pre-built agent with the actual program loaded into memory. The agents of Langchain have been used to invoke the API to generate the unit test cases. We have configured the models in two ways, one by restricting it to 2K tokens and the other one to 4K tokens. For each of the agents, we always have set the temperature to zero for the experiments to be more deterministic.

For code generation, each of the candidate test cases was first processed to generate a code generation template. The code generation templates have the following attributes present in them:
\begin{itemize}
    \item The header files and libraries are present at the beginning of the file.
    \item Detailed comment on the test case that it is supposed to test.
    \item All the global variables and helper functions are preserved.
\end{itemize}

The files are treated more as code completion task inputs rather than a complete generation from scratch. This helps to narrow down the variance by guiding the LLMs to use specific kinds of libraries to generate test functions. We do this for each test case in three different temperature settings ($0$, $0.2$, and $0.4$) and observe changes in code generation behaviour. The Langchain agents that we built can help to run the code generation task in scale in a controlled manner.

\begin{lstlisting}[language=C++,caption={A sample unit test from DBCSR test suite},label=listing:dbscr1, basicstyle=\tiny]
#include <vector>
#include <iostream>
#include <algorithm>
#include <cstdlib>
#include <cstdio>
#include <cstdint>
#include <random>
#include <mpi.h>
#include <dbcsr.h>
// Random distribution by using round-robin assignment
// of blocks to processors
std::vector<int> random_dist(int dist_size, int nbins) {
  std::vector<int> dist(dist_size);
  for (int i = 0; i < dist_size; i++) dist[i] = i % nbins;
  return dist;
}
int main(int argc, char* argv[]) {
  MPI_Init(&argc, &argv);
  int mpi_size, mpi_rank;
  MPI_Comm_size(MPI_COMM_WORLD, &mpi_size);
  MPI_Comm_rank(MPI_COMM_WORLD, &mpi_rank);
  // Make 2D grid
  ...............
  ...............
\end{lstlisting}

In Listing \ref{listing:dbscr1}, we have a sample of how this looks like for the OOB test code generation. 
This is a test from the DBCSR~\cite{dbcsr} project with the real test methods removed. However, it does have the guiding libraries and code comments present for the LLMs when generating the test code.

For each of the evaluated LLMs, we evaluated the three responses out of the thirty that were produced and included them in the analysis.

\subsection{Code Template Guided Generation}\label{contextguide}

Even though OOB generation already guided the LLMs towards using certain types of libraries and variables, it still left the LLMs up for open interpretation about what a test case would be. We also tested the models' capability in a more controlled and guided way. The goal was to investigate if such an approach would result in better test coverage.

For a template-guided generation, one can formalize the unit test generation model as a function $F$. 
Given a prompt $x$, containing the original code for which the unit test needs to be written, the LLM can complete $x$ to get $y=F(x)$, where $y$ is the response to the provided prompt $x$. 
This makes the entire code prompt (input prompt containing source code along with code template from memory together with the response) as a grouping of $A_1 + A_2$, where $A_1$ is the input prompt $x$, and $A_2$ is the output $y$ given $A_1$.

We can generate multiple outputs (completions) by using different sampling strategies: Random sampling and Beam search. Random sampling is a simple approach that selects the next token at random, and Beam search is a more sophisticated approach that takes into account the probability of each token~\cite{deshpande2019fast,wang2017diverse}.

\subsubsection{Contextual Code Generation}

We created the prompts as part of the Langchain agents and ran them 10 times with new memory to generate 10 separate responses for a specific method. 
Each prompt pattern follows a similar format to classic software patterns, but with slight modifications to match the context of output generation with the LLMs. It includes a name and classification, the intent and context, the motivation, the structure and key ideas, an example implementation, and consequences. The prompt pattern name uniquely identifies the pattern and indicates the problem being addressed, while the classification includes categories of pattern types such as output customization, error identification, prompt improvement, interaction, and context control. We created a Langchain memory using the context that encompasses the source code and also the libraries, global variables and classes from the original unit tests to guide the resulting output in a better way.

Listing~\ref{listing:optim}  illustrates the structure of these prompts and context, in which lines 1-24 and 28-36 are part of the context and prompt.

\begin{lstlisting}[language=C++,caption={Testcase generation template},label=listing:optim, basicstyle=\tiny]
// SUMT test
#include "optim.hpp"
#include "./../test_fns/test_fns.hpp"
int main()
{
    //
    ColVec_t x_1 = BMO_MATOPS_ONE_COLVEC(2);
    bool success_1 = optim::sumt(x_1,constr_test_objfn_1, nullptr,constr_test_constrfn_1, nullptr);
    if (success_1) {
        std::cout << "sumt: test_1 completed successfully." << std::endl;
    } else {
        std::cout << "sumt: test_1 completed unsuccessfully." << std::endl;
    }
    BMO_MATOPS_COUT << "sumt: solution to test_1:\n" << x_1 << "\n";
    //
    ColVec_t x_2 = BMO_MATOPS_ONE_COLVEC(2);
.............
.............
.............
    // coverage tests
    optim::algo_settings_t settings;

    optim::sumt(x_1,constr_test_objfn_1, nullptr,constr_test_constrfn_1,nullptr);
    optim::sumt(x_1,constr_test_objfn_1, nullptr,constr_test_constrfn_1,nullptr, settings);
    return 0;
}
\end{lstlisting}

\begin{figure}
    \centering
    \includegraphics[scale=0.35]{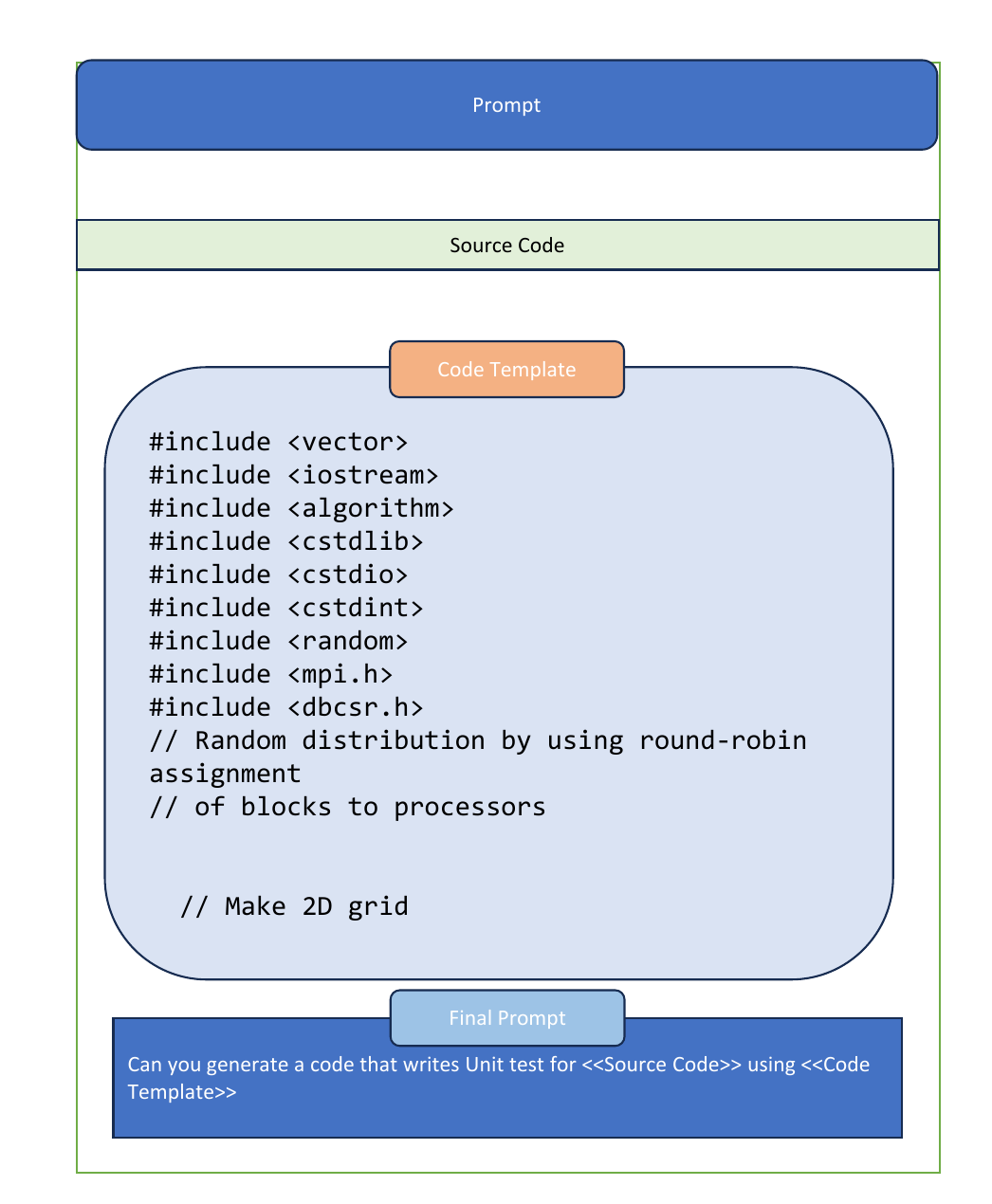}
    \caption{An example of generating unit test using prompt.}
    \label{fig:prompteng}
\end{figure}

\subsubsection{Analysis} 

The code generation can be illustrated as depicted in \figurename~\ref{fig:prompteng}.
For analysis of the generated unit tests, we look for code compilability, code correctness, and parallelism detection on the unit tests. For our evaluation, it is imperative that we use the existing human-written unit tests as the gold standard and compare them with the generated ones.


\subsubsection{Compilation Test}

We used the existing makefiles of the projects to take the generated unit tests as input as a plug-and-play method so that the existing infrastructure was leveraged to evaluate whether the unit test succeeds or not. 
The test code which did not succeed was then inspected to prune out any kind of non-coding error that might have prevented the test from successful execution. We observed that some of the compilation errors were caused by syntax problems that could be fixed by a heuristic-based post-processing step after unit test generation. More specifically, we have noticed certain patterns in the generated test cases: 

\begin{itemize}
\item Generate an extra test class that is incomplete. Thus producing unused code.
\item Include natural language explanations before and/or after the code. This was primarily noticed in code completion-based tasks, where we were guiding the output to become a code-infilling task.
\item Repeat the code under test and/or the prompt.
\item Change libraries and declarations or at times remove them.
\item Generate incomplete unit tests after it reaches its token size limit.
\end{itemize}

We mitigated some of these by post-processing the code with a rule-based fixer that we have created. For the token size limitation, we attempted to utilize Langchain memory to pass on the truncated part of the code to send as a reference to generate the remaining part, essentially making it a code completion task. For code comments being introduced, we removed them as much as possible using rule-based heuristics. This was more prevalent in the Davinci~\cite{godoy2023evaluation} model.

\section{Evaluation}

For evaluation of the generated test code. We ran each test case through a series of tests, each designed to check different aspects of the generated code. The first evaluation was to find out how many of them can be compiled through the OOB method, and how many of those that have failed can be ``fixed'' using the contextual inputs. The series of tests that have been applied to evaluate the generated code are described in the following subsections. 

\subsection{Test Coverage}
We evaluated the line coverage and branch coverage of the generated unit tests and compared them with the original human-written unit tests from our selected project repositories in \tablename~\ref{projects}. 
\begin{itemize}
    \item \textbf{Line Coverage:} In line coverage, we measure how many lines of the code from the original source code have been executed by the unit test out of the total number of lines present in the code.
\begin{equation}
\text{Line Coverage} = \frac{{\text{{No. of lines executed}}}}{{\text{{No. of lines in source code}}}} \times 100
\end{equation}

\item \textbf{Branch Coverage}: It is a software testing metric that measures how many branches in a program that are executed by a test suite. It is calculated by dividing the number of branches executed by the total number of branches in a program. A branch is a point in the program where the control flow can go in one of two or more directions. For example, an if statement has two branches: one for the true condition and one for the false condition.

\begin{equation}
\text{Branch Coverage} = \frac{\text{No. of branches executed}}{\text{Total no. of branches}} \times 100
\end{equation}
\item Assuming the code has been implemented correctly, we evaluated whether the generated unit tests check the correct functions for correct outputs. 
\end{itemize}

For each of these, we used certain pre-defined parameters as our indicators. For the line and branch coverages, we used the open-source tool OpenCppCoverage \cite{OpenCppCoverage}, which provides the information needed for coverage evaluation.

\subsection{RQ1: Parallel and High Performance Program Unit Test Generation using the LLMS}

We analyzed the generated unit tests for parallel and high performance programs according to four dimensions following Siddiq et al.~\cite{siddiq2023exploring}:
\begin{itemize}
\item \textbf{Compilation Status}: We assessed whether the generated unit tests could be compiled without errors.
\item \textbf{Correctness}: We examined whether the generated unit tests could pass when running against the code under test.
\item \textbf{Coverage}: We checked whether the generated unit tests covered all of the code under test.
\item \textbf{Quality}: We evaluated whether the generated unit tests were of high quality, in terms of factors such as readability, maintainability, and testability.
\end{itemize}

Apart from these above, we also investigated whether we could introduce additional parameters into the unit tests that were not originally covered by the existing test. Specifically, these include: i) Hierarchical Parallelism and ii) Mathematical Functions.
Hierarchical parallelism is a type of parallelism that uses multiple levels of parallelism to improve performance. For example, a program might use thread-level parallelism to parallelize loops within a process, and process-level parallelism to parallelize multiple processes.

\subsection{Compilation Status}

Since our code repositories were carefully chosen to have good quality unit tests, we could utilize the Makefiles for each project. We used these Makefiles to just change the target unit tests from human written ones to the LLM generated ones for our compilation check. This also helped us to log compilation status for successful and unsuccessful compiles for post processing and data analysis.
As shown in the second column of Table \ref{llmtest}, less than 50\% of the generated unit tests are compilable across all the studied models. Among the compilable tests, the ratio was noticeably lower when we just tried to generate using the partial test case and the name as input to the model.

\begin{table}[h]
\centering
\footnotesize
\caption{Compilation status of the generated unit tests}
\label{llmtest}
\label{tab:compilation-status}
\begin{tabular}{lccccc}
\hline
\textbf{LLM} & \textbf{\% Compilable} & \textbf{\% Context Aware} \\
\hline
ChatGPT 3.5 & 23.1\% & 61.3\%  \\
Davinci (2K) & 27.5\% & 80\%\\
Davinci (4K) & 24.4\% & 79.4\% \\
Manual & 100\% & NA \\
\hline
\end{tabular}
\end{table}

After applying the context-aware method described in Section \ref{contextguide}, we observed that we were able to automatically fix several unit tests (as shown in Table \ref{llmtest}). For ChatGPT, the increase is from 23.1\% to 61.3\%. For Davinci, it's almost 80\%. The results raise a very interesting observation as ChatGPT was using gpt-3.5-turbo~\cite{openai2022chatgpt} which should be the superior model for code generation among the three. However, our code-infilling approach has produced worse results than a prompt-based regeneration.

Since these were done to the code that we already fixed using post-processing described earlier in section \ref{testcgen}, we clustered the remaining compilation error using the K-Means algorithm similar to Siddiq et al.~\cite{siddiq2023exploring}. We used the Elbow Rule to find out the optimal K with the objective of identifying the specific compilation errors that are prevalent for the parallel and high performance programs.
In this case, the K-means algorithm was able to cluster the compilation errors into 41 different groups. This might suggest that there could be 41 different types of semantic errors that may cause compilation failures. The Silhouette method showed that the clustering solution was of good quality, which means that the clusters are well-separated from each other. We have made the observation that
the top compilation errors for the dataset were attributed to the
missing OpenMP pragmas (OpenMP pragmas are used to control the execution of OpenMP code. If these pragmas are missing, the compiler will not be able to compile the code),
and incorrect OpenMP pragmas (OpenMP pragmas must be used correctly. If they are not, the compiler will not be able to compile the code). Among the errors caused by
incorrect method invocations, 51\% of them were invocations
to an assertion method ``CPPUNIT\_ASSERT\_MESSAGE''.

\subsection{Test Correctness}

We used the following two metrics for measuring the correctness of a unit test, and evaluation results are reported in the \tablename~\ref{tcorrect}.
\begin{itemize}
    \item \textbf{Fully Correct}: The unit test has a success rate of 100\%, meaning that all of its test methods have passed.
    \item \textbf{Somewhat Correct:} The unit test has at least one passing test method.
\end{itemize}

\begin{table}[htbp]
\centering
\footnotesize
\caption{Percentage of Correct Tests}
\label{tcorrect}
\begin{tabular}{p{1.1in}p{0.5in}p{0.5in}p{0.5in}}
\toprule
\textbf{Model} & \textbf{ChatGPT-3.5} & \textbf{Davinci (2K)} & \textbf{Davinci (4K)} \\
\midrule
\textbf{Fully Correct} & 22.3\% & 47.5\% & 46.7\% \\
\textbf{Somewhat Correct} & 62.3\% & 57.5\% & 57.4\% \\
\midrule
\end{tabular}
\end{table}

These results from \tablename~\ref{tcorrect} show that even though we could not generate correct unit tests through any of the models on a large scale, they are still useful for generating some correct ones. We also noticed that increasing code context by using the Langchain memory did not help with the numbers. However, we observed that there is an increase in the range of 2\% when adding compiler-guided feedback into the memory for subsequent runs.
In other words, the models were able to generate unit tests that could pass at least one test method, even if they were not able to generate unit tests that could pass all of the test methods.
It was observed that increasing the size of the token vocabulary did not improve the correctness of the generated unit tests. This suggests that the size of the token vocabulary is not the only factor that affects the correctness of the generated unit tests. Other factors, such as the quality of the training data, may likely play a role.

\subsection{Test Coverage}

We assessed the line coverage and branch coverage of the generated unit tests and compared the results with the coverage of the original tests. We used gcov~\cite{gcov} to measure line and branch coverage. Line and branch coverage are two metrics used to measure the effectiveness of unit tests. Line coverage measures the percentage of lines of code that are executed by the unit tests. Branch coverage measures the percentage of branches in the code that are executed by the unit tests.
The results have been highlighted below in \tablename~\ref{testcov}.

\begin{table}[htbp]
\centering
\footnotesize
\caption{Line and Branch Coverage}
\label{testcov}
\begin{tabular}{p{1in}p{0.5in}p{0.5in}p{0.5in}}
\toprule
\textbf{Metric} & \textbf{ChatGPT-3.5} & \textbf{Davinci (2K)} & \textbf{Davinci (4K)} \\
\midrule
\textbf{Line Coverage} & 59.1\% & 77.4\% & 77.7\% \\
\textbf{Branch Coverage} & 56.5\% & 82.1\% & 82.8\% \\
\bottomrule
\end{tabular}
\end{table}

In Table \ref{testcov}, the line coverage and branch coverage for each model are shown. For example, ChatGPT-3.5 has a line coverage of 59.1\% and a branch coverage of 56.5\%. This means that 59.1\% of the lines of code in the code under test were executed by the unit tests from ChatGPT-3.5, and 56.5\% of the branches in the code under test were executed by the unit tests from ChatGPT-3.5.

In general, higher line and branch coverage is considered to be a good property, as it indicates that the unit tests are effective in exercising the code under test. However, it is important to note that high line and branch coverage does not guarantee that the unit tests are complete or effective. Other factors, such as the quality of the unit tests, also play a role in determining the effectiveness of unit tests.

\subsection{Quality}

We used test smells as the indicator for code quality. Test smells can be applied for defining quality because they can indicate potential problems with the unit tests. These problems can make it difficult to maintain the tests, and they can also lead to false positives and negatives. As a result, it is important to identify and address test smells as early as possible.

Test smells are code smells that can indicate problems with unit tests. They can be caused by a variety of factors, such as poor design, bad practices, or simply a lack of attention to details.
Following Siddiq et al.~\cite{siddiq2023exploring}, we studied the following test smells in the generated parallel program unit tests by the LLMs:
\begin{itemize}
    \item \textbf{Assertion Roulette (AR)}: This occurs when the unit test randomly selects values to assert against.
    \item \textbf{Conditional Logic Test (CLT)}: This occurs when the unit test checks the output of a conditional statement.
    \item \textbf{Constructor Initialization (CI)}: This occurs when the unit test checks the output of a constructor.
    \item \textbf{Empty Test (EM)}: This occurs when the unit test does not actually test anything.
    \item \textbf{Exception Handling (EH)}: This occurs when the unit test checks the output of exception handling code.
    \item \textbf{Redundant Print (RP)}: This occurs when the unit test prints the same output multiple times.
    \item \textbf{Redundant Assertion (RA)}: This occurs when the unit test asserts the same thing multiple times.
    \item \textbf{Sensitive Equality (SE)}: This occurs when the unit test uses sensitive equality operators, such as == or !=, to compare values.
    \item \textbf{Sleepy Test (ST)}: This occurs when the unit test takes a long time to run.
    \item \textbf{Eager Test (EA)}: This occurs when the unit test asserts the output of code that is not yet executed.
    \item \textbf{Lazy Test (LT)}: This occurs when multiple test methods invoke the same production code.
    \item \textbf{Duplicate Assert (DA)}: This occurs when the same assertion is repeated multiple times in the same test method.
    \item \textbf{Unknown Test (UR)}: This occurs when the unit test does not actually test the code that it is supposed to test.
    \item \textbf{Ignored Test (IT)}: This occurs when the unit test is marked as ignored.
    \item \textbf{Magic Number Test (MNT)}: This occurs when the unit test hard-codes a value in an assertion without a comment explanation.
\end{itemize}

From the results of Table~\ref{testsmells}, we can observe that the most prevalent smell types were Magic Number Test (MNT) and Lazy Test (LT) across all the approaches, i.e., in the unit tests generated by the LLMs, 
The MNT smell occurs when the unit test hard-codes a value in an assertion without a comment explaining it. This can make it difficult to understand what the unit test is testing and can also make it difficult to maintain the unit test.
The LT smell arises when multiple test methods invoke the same production code. This can make it difficult to determine which test method is responsible for a failure and can also make it difficult to refactor the production code.

\begin{table}
\centering
\footnotesize
\caption{Test Smells Distribution}
\label{testsmells}
\begin{tblr}{
  width = \linewidth,
  colspec = {Q[115]Q[212]Q[212]Q[237]Q[144]},
  column{even} = {c},
  column{3} = {c},
  vlines,
  hline{1-2,17} = {-}{},
  hline{3-16} = {5}{},
}
\textbf{Smell} & \textbf{Davinci (2K)} & \textbf{Davinci (4K)} & \textbf{ChatGPT-3.5} & \textbf{Manual} \\
AR             & 75.3\%              & 69.7\%              & 33.8\%               & 0.0\%           \\
CLT            & 0.0\%               & 0.0\%               & 4.5\%                & 0.0\%           \\
CI             & 0.0\%               & 0.0\%               & 0.0\%                & 0.0\%           \\
EM             & 5.9\%               & 1.3\%               & 0.8\%                & 0.0\%           \\
EH             & 0.0\%               & 0.0\%               & 0.0\%                & 100.0\%         \\
RP             & 0.0\%               & 0.0\%               & 0.0\%                & 0.0\%           \\
RA             & 0.0\%               & 0.0\%               & 0.0\%                & 0.0\%           \\
SE             & 0.0\%               & 0.0\%               & 0.0\%                & 0.0\%           \\
ST             & 0.0\%               & 0.0\%               & 0.0\%                & 0.0\%           \\
EA             & 60.9\%              & 59.2\%              & 33.8\%               & 16.3\%          \\
LT             & 49.4\%              & 51.4\%              & 96.2\%               & 99.4\%          \\
DA             & 25.6\%              & 34.5\%              & 4.1\%                & 0.6\%           \\
UT             & 10.0\%              & 7.7\%               & 1.8\%                & 0.0\%           \\
IT             & 0.0\%               & 0.0\%               & 0.0\%                & 0.0\%           \\
MNT            & 100.0\%             & 100.0\%             & 100.0\%              & 100.0\%         
\end{tblr}
\end{table}

Moreover, Table \ref{testsmells} provides an overview of the performance of all the models in this case. The most prevalent test smell is Assertion Roulette (AR), which is present in 75.3\% of the unit tests generated by Davinci (2K), 69.7\% of the unit tests generated by Davinci (4K), and 33.8\% of the unit tests generated by ChatGPT-3.5. The next most prevalent test smell is Lazy Test (LT), which is present in 96.2\% of the unit tests generated by ChatGPT-3.5, 51.4\% of the unit tests generated by Davinci (4K), and 99.4\% of the unit tests in the GitHub Repository (Manual). Other notable test smells include Conditional Logic Testing, Concurrent Invocation, Empty Method, Exception Handling, Redundant Assertion, Resource Allocation, System Execution, Sleepy Test, Excessive Assertion, Default Assertions, and Unreachable Test. These results can be used to improve the quality of unit tests by identifying and addressing test smells.

\subsection{Parallelism}

To evaluate whether the generated parallel and high performance program unit tests cover the original source code for parallelism we ensured that they have two categories of tests: \textbf{1)} memory copy, reduction, and atomic, 
and \textbf{2)} Multiple datatypes: float, double$<$complex$>$.

Tests should be \textbf{i.} Self-contained,
\textbf{ii.} Return 0 on success and not 0 on failure, and 
\textbf{iii.} Independent of the number of teams/threads.
We used open source tool OvO\cite{ovo} to compare the results of our LLM-generated unit tests to see if they match.

\section{Summary \& Discussion}

In this section, we summarize the results of our studies and discuss the implications of the findings.

Writing unit tests for OpenMP and MPI codes is distinctively different from testing regular C++ code due to the inherent parallel and distributed nature of these frameworks. With standard C++ code, one typically deals with a sequential program flow, and the tests are designed to evaluate specific functionalities in that flow. However, with OpenMP (for shared-memory parallelism) and MPI (for distributed-memory parallelism), the challenge not only includes ensuring functional correctness but also assessing the proper synchronization, data sharing, and communication among threads or processes. For OpenMP, tests need to check for race conditions, deadlocks, and proper handling of shared and private variables. In the MPI context, the challenges lie in testing message passing correctness, ensuring that data is communicated correctly between processes, and managing deadlocks due to unmatched sends and receives. Validation involves checking that the code scales correctly with increasing number of threads or processes, that load balancing is achieved if applicable, and that there is no unintended data leakage or corruption across the parallel regions or during inter-process communications. All these make it challenging for LLMs trained on general purpose code to properly encode the knowledge needed to write test cases for HPC code.

In our work we have noticed that \textbf{the LLMs can generate correct unit tests} both without any guidance and with guidance for the parallel code. Often however the test code did not explicitly check parallelism. So even though the tests passed the \textit{correctness} check, for functional purposes, they are inferior to the human written tests.

\textbf{LLMs tend to generate better unit tests} and increase code coverage when provided with the source code as context and code template as guidance. Essentially reducing the task to be a code infilling task. 

\textbf{LLMs may not take into account libraries} often leading to missed library or variable name inclusion in the actual code but not having correct reference, thus making the code failed to compile.

\textbf{LLMs have a tendency to introduce non code segments} into the code as well as unit tests. These mostly were removed using a post processing step during our evaluation.

\textbf{LLMs tend to hallucinate when provided with ``too much'' context}. One interesting observation was that when we have built our agents and memories using Langchain, often sending larger source code and code templates as guidance resulted in getting an output code with nonexistent types methods etc.

The Large language models (LLMs) often produce code that includes non-existent types, methods, and other constructs, which can result in compilation errors and related issues. For instance, in the case of Codex, it generates inputs with types like Tuple, Pair, Triple, Quad, and Quint that do not exist.

These code artifacts are not specific to high performance software or unit tests though. However the code correctness improvements using template based guided approach shows that the approach can be a good way to improve code synthesis for unit test case generations. Since unite tests are not self-contained code generation task, rather a specific task based on pre-defined existing code, we can use code interpretation techniques to come up with better guidance and enhancement to improve the quality of the generated test cases.

\subsection{Concerns}

In our investigation, We have used a version of Davinci (text-davinci-002) that was trained on data up to September 2021. This means that the tests in our benchmarks were also part of its training set. This raises the possibility that the model may be memorizing the existing tests, rather than generating new ones. This could limit the model's usefulness for code projects that it was not trained on.

One approach to investigate the potential effects of memorization is to measure similarity through edit distance between the generated code and the existing one. Prior work has shown that code plagiarism~\cite{lemieux2023codamosa} or clone detection~\cite{schleimer2003winnowing} techniques might not be effective at identifying LLM code memorization.

Edit distance measures the dissimilarity between two strings by tallying the edits required to align one string with the other. We employed this metric to gauge the resemblance between each test produced and the pre-existing tests in our benchmarks. Our observations indicated that models tended to replicate existing tests when the newly generated test necessitated only a few modifications to match the benchmark test. 

\section{Threats to Validity}

Although our evaluation involved a comprehensive scale that surpasses the previous test generation approaches such as \cite{bareiss2022code}, it is important to note that our results are derived from an analysis of 7 parellel and high performance projects, which may have limitation in terms of generalizability to other high performance parallel program codebases. To address this concern, we have taken several measures: (i) diversifying the parallel code repos to incorporate different kinds of hardware used in those and different domains; (ii) choosing newer test cases that should not have been present in the original training dataset of the LLMs; and (iii) assessing the similarity between the generated tests and the existing tests. Despite these improvement efforts, it is crucial to recognize the potential limitation and consider further investigations across a wider range of codebases to achieve better generalizability. 

\section{Related Work}

The related work can be broadly categorized into the following areas. 

\paragraph{Research in parallel and high performance software unit testing}  Although TDD (Test-driven development) and unit testing framework are well established practice in software engineering, adopting them in the context of high-performance computing environments often faces unique requirements and challenges. High performance computing developers may encounter problems that differ from other application areas of programming~\cite{8452053}. For instance, a specific need in parallel and high performance computing is to develop tools for supporting unit tests that can be integrated with certain legacy HPC code~\cite{7839465}. Another effort is to create unit testing framework tailored for the HPC environments and applications, for instance extension of unit testing framework to support parallelization techniques in high performance applications such as OpenMP and MPI.  

\paragraph{Evaluation of the LLMs in code generation} Another area of related work is evaluation of the LLMs for code generation~\cite{chen2021codex, 10.1145/3520312.3534862}. A particular topic that has attracted large amount of attention from the research community is to evaluate the code quality and security aspect of the LLM generated code~\cite{pearce2021asleep, perry2022users}. For instance, in~\cite{10.1145/3558489.3559072}, the authors have assessed the quality of the LLM generated code from multiple aspects such as compilation, functional correctness and code efficiency. Researchers also conducted studies of the bugs in the LLM generated code, which shows that there is a high chance that the LLMs may introduce difficult to detect bugs in the generated code~\cite{10174227}. Besides bugs in the LLM generated code, researchers have scrutinized and assessed the code generated by the LLMs for security vulnerabilities~\cite{pearce2021asleep}. Quality of the LLM generated code is also investigated from the reliability and robustness perspective~\cite{zhong2023study}. 

\paragraph{Application of the LLMs for unit testing} Last but not the least, researchers have explored and evaluated applicability of the LLMs for unit test generation~\cite{yuan2023manual}. For instance, Bareiß et al.~\cite{bareiss2022code} examined the performance of Codex in Java code test generation. They applied a few-shot learning approach, where the model was provided with a function to be tested, an example of another function, and an associated test. This allowed the model to learn the expected structure of a test. In another study~\cite{tufano2020unit}, the authors evaluated test coverage of a BART transformer model fine-tuned on a training dataset consisting of functions and tests. 

High-Performance Computing (HPC) has become a pivotal tool for scientific and engineering advancements, enabling the resolution of intricate problems that demand substantial computational prowess \cite{dambra2003}. As these challenges intensify, the imperative for robust and efficient software solutions becomes evident. This realization has spurred a growing inclination towards the incorporation of established software engineering practices, such as Test-Driven Development (TDD) and unit testing, within the high performance realm \cite{rilee2014}. However, the transition isn't straightforward. The unique requirements of parallel and high performance domain, like managing vast datasets, safeguarding data integrity across distributed systems, and optimizing for performance on dedicated hardware, pose distinct challenges \cite{cartier2013}. 


This work distinguishes from all the related work with its focus on high performance related applications and parallel code. To the best of our knowledge, this is the first step investigating and evaluating the LLMs in the context of parallel and high performance programming and testing. Although the study only scratches the surface of this potentially rich area of endeavor, it shows initial results that could pave the way for further research and investigation. 

\section{Conclusion}

Testing parallel and high performance applications is often considered to be difficult. Prior efforts in parallel program testing have focused on generating test cases for legacy high performance code, testing support for parallel programs. Manually crafted high performance computing test cases are not only laborious to develop but also time-consuming. The recent advances of code-generating Large Language Models (LLMs) have the potential to revolutionize the field of software testing. The described study has filled a research gap to investigate the feasibility and effectiveness of applying the generative models (e.g., Davinci, and ChatGPT) for generating C++ parallel program test cases. We have reported the early discoveries that show potential benefits in coverage and test smell.  The study paves a road for further exploration in this exciting and important new research direction. 

\bibliographystyle{unsrtnat}
\bibliography{references}
\end{document}